# A deep learning based solution for imperfect CSI problem in correlated FSO communication channel

M. A. Amirabadi

Email: m_amirabadi@elec.iust.ac.ir

**Abstract-** Imperfect channel state information (CSI) at the receiver, which is due to channel estimation error, is one of the main problems toward achieving optimum detection. This paper presents a deep learning based structure for combating this issue. In order to show the effect of using deep learning, the symbol error rate of a simple free space optical (FSO) communication system is simulated over correlated and un-correlated log-normal channel with write/ wrong CSI. Novelties and contributions of this paper, which are done for the first in machine learning for FSO communication include considering deep learning, considering Log-normal channel, considering correlated channel, considering imperfect CSI. The proposed deep learning based structure is compared with maximum likelihood detector, it is shown that in perfect CSI, both perform the same (because maximum likelihood is optimum), but in imperfect CSI, proposed deep learning based structure outperforms maximum likelihood in channels with un-correlation or desired correlation lengths.

## I- Introduction

Wireless optical communication is the main competitor of the conventional radio frequency systems [1]. In indoor the visible light communication, in outdoor the free space optical (FSO) communication, and in underwater optical communication have shown great performance, data rate, security [2]. This fact should also be mentioned that their installation is easier and there are low cost. The requirements of the next generation optical communication systems might lead to complete replacement of the conventional and model wireless communication system [3]. For this purpose, it is highly required to investigate these systems over varies system and channel scenarios.

Considering FSO channel effects, weather condition, pointing error, as well as atmospheric turbulence are the most important [4]. Even in clear weather, short range link, with properly fastened transceiver at the top of high buildings, the atmospheric turbulence effects could not be resisted [5]. Accordingly many investigations considered various stochastic distributions to model this effect. Among these models Log-normal is best suit for weak atmospheric turbulence regimes [6, 7]. However, mostly in investigations the FSO channel is considered un-correlated, in the sense that it is assumed that the sampling is done such that the sampling period is longer than the channel correlation time. Accordingly, it might be more interesting to see what happens when this thing does not be applied. Another issue that is hardly ever investigated, is the effect of imperfect channel state information (CSI) due to the channel estimation error. However, one of the main drawbacks toward conventional systems in considering these effects (correlated channel and imperfect CSI) is the complexity of the receiver. Because usually the FSO receivers are implemented in urban environments (despite fiber that is used in the links between urban areas); and in these areas users are the common costumers; so the cost, complexity and performance should be deserved at once to gather.

Recently, a new hot topic emerged in optical communication entitled machine learning; at first should be told that machine learning is exactly for problems with complex conventional solution. Despite all conventional techniques, in machine learning it is assumed that the solution is known at the training phase, and this really could help the machine to do properly in testing phase. However, it is completely dependent on the type of the machine, on the amount and type of input features of the machine, on the tuning of hyperparameters. Common machine learning techniques are proper for low complexity solutions. For problems with higher complexity, deep learning is more preferred to machine learning. Deep learning, which is actually an extension of machine learning, is able to drive very complicated relationships between input features and desired output, of course with the cost of more complexity. One of the mostly used types of deep learning algorithms is deep neural network (DNN). DNN is actually an extended form of artificial neural network (ANN), which its idea and structure is taken from the human brain neural network. In this network several layers, which each is composed of many neurons are connected together. The output signal of one neuron is multiplied by a weight, and entered the next neuron. The entered signals to a neuron are summed, added by a bias and passed through a so-called activation function. The desired output and the actual output of the DNN are feed into a cost function, the aim of training is to adjust weight and biases such that the cost function be minimized.

Considering DNN applications in optical communications, many investigations are done, which are mostly over fiber optical communication. The applications such as fiber linear and nonlinear effects mitigation [8], optical amplifier control [9], modulation format identification [10], optical performance/ power monitoring [11], etc. However, in wireless optical communication, and especially is FSO communication, there is a lack of DNN application. It should be noted that not only DNN, but also other machine learning and deep learning algorithms are rarely used in FSO communication, in the following, the investigations over machine/deep learning for FSO communication are reviewed. Considering machine learning applications in FSO, few basic investigations are developed in applications such as detection [12], distortion correction for sensor-less adaptive optics [13], and demodulator for a turbo-coded orbital angular momentum [14]. All of these works considered machine at the receiver side of a simple FSO-SISO system for a limited (specific) scenario; there is lack of comprehensive investigation in machine /deep learning for FSO communication.

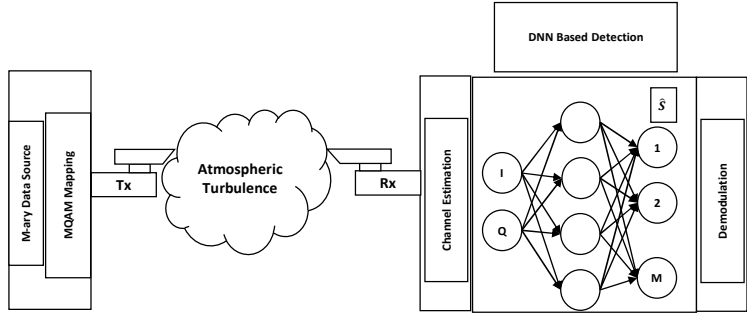

Fig.1. proposed DNN based system model.

According to the above discussions, mostly in FSO communication applications, ideal signal processing and channel modeling scenarios are considered to reduce the complexity of the problem solution; on the other hand deep learning is one of the low complex solutions that could afford non-linear real system/channel problems. So, it might be interesting to consider a more realistic scenario such as correlated channel with imperfect CSI, and use deep learning as a detector to solve this problem and compare it with the conventional techniques from performance and complexity view. Accordingly, this paper considers a simple FSO system and deploys the mentioned scenario on it. The symbol error rate (SER) of this system is then investigated in Log-normal atmospheric turbulence regime. It is expected that deep learning could outperform conventional techniques, because actually in training phase, DNN has the solution. To the best of the author's knowledge, the contributions and novelties of this work, which are done for the first time in machine learning for FSO communication include the following;

1- Considering DNN,
2- Considering Log-normal channel model,
3- Considering imperfect CSI,
4- Considering correlated channel.

The rest of this paper is organized as follows; section II describes the system model, section III is the results and discussions, section IV is the conclusion of this work.

II- **System model**

The proposed DNN based system model is shown in Fig.1. Considering $x$ as the generated signal, it is first converted to a one hot vector, then mapped to an M-ary QAM constellation. Then the mapped signal and added by a unit value DC bias (in order to be positive) and then converted to FSO signal and is transmitted through a coherent optical transmitter, encountered by correlated/ un-correlated Log-normal atmospheric turbulence, and added by additive white Gaussian noise (AWGN) at the receiver. The received optical signal is converted to electrical signal and the DC bias is removed. Considering perfect/ imperfect CSI, the complex electrical signal is fed into a DNN with 2 input neurons, $M$ output neurons, $N_{hid}$ hidden layers, $N_{neu}$ per hidden layer neurons, $\alpha(.)$ Activation function, $\boldsymbol{W}$ weight matrix, and $\boldsymbol{b}$ bias vector. The aim is to train the DNN (adjust $\boldsymbol{W}$ and $\boldsymbol{b}$) such that the actual output of the DNN be equal to the desired output (which is the transmitted one hot vector). To this aim a proper loss function should be chosen; usually MMSE or cross entropy are used for DNN applications in optical communication. Considering the actual output as $\hat{\boldsymbol{s}}$, and desired output as $\boldsymbol{s}$, the cross entropy loss function would be $L(\boldsymbol{s}, \hat{\boldsymbol{s}}) = \frac{1}{K}\sum_{k=0}^{K}\sum_{i=0}^{M} s_{i,k} \log(\hat{s}_{i,k})$, where $K$ is the batch number. Actually, the aim is to minimize this loss function, to this aim various optimization methods could be used; one of the most promising is the stochastic gradient descent method. Among gradient descent methods, Adam is the mostly used and showed better results due to the mathematical manipulations which led to it [15].

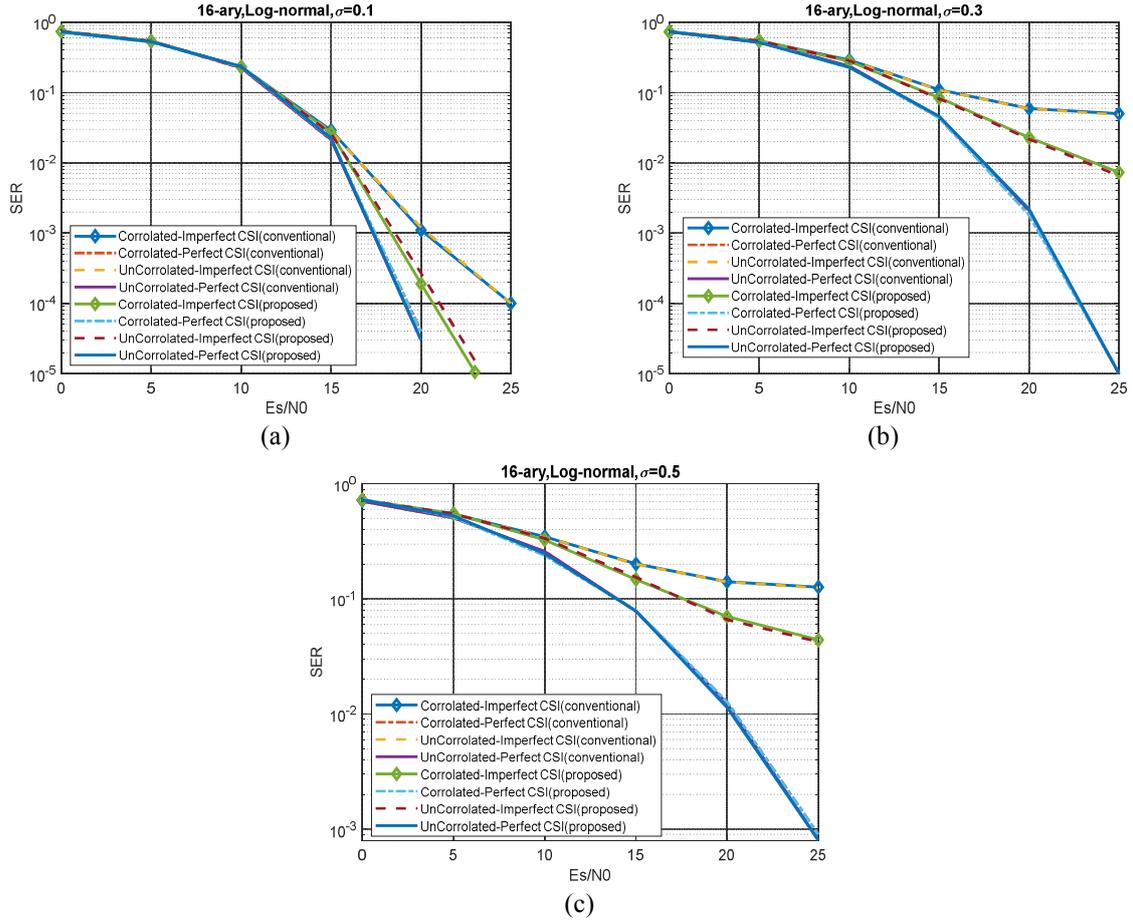

Fig.2. SER of the proposed system as a function of Es/N0 for a. $\sigma = 0.1$, b. $\sigma = 0.3$, c. $\sigma = 0.5$, when modulation order is $M = 16$, and correlation length is 2.

Actually minimization of a loss function leads to a good result when the hyperparameters be tuned properly. The difference between a well-tuned DNN and a non-tuned DNN, is the difference between a machine that could learn complex relationships and a machine that could not learn anything. So, tuning hyperparameters such as batch size, number of batches, learning rate, number of iterations, number of neurons, number of hidden layers, activation function, etc., are very important in training a DNN. There are different ways for this aim which [15] properly investigated over this topic and is very useful.

### III- **Results and Discussions**

In this section simulation results of the proposed deep learning based structure are presented, and compared with conventional structures. Deep learning based structures are simulated using Python/ Tensorflow environment, because it is super helpful for this aim, and simulations of conventional systems are done with Matlab. The proposed FSO system is considered at correlated/ un-correlated Log-normal atmospheric turbulence with perfect/ imperfect CSI. Different Log-normal variances, as well as different correlations are considered. The selection of hyperparameters is done roughly by manually selecting some parameters and observing the results. Considering sample size to batch size ratio equal to 4 has better performance in tuning. The layer types are deep artificial neural networks, which are selected due to the application type, this type can approximate any arbitrary, nonlinear, continuous, multidimensional function. There are many activation functions available in Tensorflow, e.g., sigmoid, and hyperbolic tangent, which are bounded, continuous, monotonic and continuously differentiable. However, these functions contemplate all neurons and have high complexity. Elu community such as Relu, Relu6, Crelu, and Elu have lower complexity. In hyperparameter tuning, Relu showed better performances. The number of hidden neurons depends highly on the degree of nonlinearity and model dimensionality of the model. Highly nonlinear systems require more neurons, while smoother systems require fewer neurons. Tuning indicates that $N_{hid} = 40$ neuron in each hidden layer has better performance. After designing the DNN, it is turn to select a loss function, MMSE, and Cross entropy are the well-known loss functions, which Cross entropy is mostly used in DNN for OC applications, and therefore is used in this paper. The last step it choosing the optimizer; among the optimizers available in Tensorflow Adam is the mostly used one and shows better performance in tuning procedure. Tuning indicates that iteration number of 1000 could bring the required accuracy.

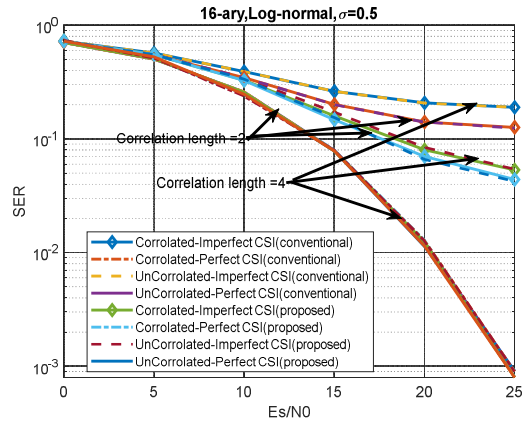

Fig.3. SER of the proposed system as a function of Es/N0 for correlation length of 2 and 4, when $\sigma = 0.5$, modulation order is $M = 16$.

In Fig.2. SER of the proposed system is plotted as a function of Es/N0 for a. $\sigma = 0.1$, b. $\sigma = 0.3$, c. $\sigma = 0.5$, when modulation order is $M = 16$, and correlation length is 2. The aim is to figure out the conventional/ proposed system performances in correlated/ uncorrelated channel with perfect/ imperfect CSI. Conventional system is composed of M-QAM modulator, and maximum likelihood detector. As could be seen, in the case of perfect CSI, conventional and proposed system performances in correlated and uncorrelated channels are almost the same, the reason is that in these cases, the optimum detector is available in the conventional system (maximum likelihood), and deep learning could not perform better than optimum (in addition it shows that the deep learning have learnt best its work). However, the benefit of deep learning appears when CSI is imperfect, e.g. when channel is correlated/ uncorrelated but obtained CSI does not considered correlation/ un-correlation. Actually, deep learning is shown in all its optical communication tasks to be a promising approach for anywhere that model is nonlinear (that is hard to obtain), or model is not available at all. As could be seen, in all variances of Log-normal atmospheric turbulence, proposed scheme could better perform than the conventional system. As could be seen, the system performances at correlated and uncorrelated channels are almost the same, this is because of considering weak atmospheric turbulence regimes (Log-normal distribution) in simulations. Actually (in deep learning case) this could be counted as one of the benefits, because there would be no need to train the system separately for correlated or uncorrelated schemes; the training could be done one time, and be used in any desired channel.

In Fig.3. SER of the proposed system is plotted as a function of Es/N0 for correlation length of 2 and 4, when $\sigma = 0.5$, modulation order is $M = 16$. As can be seen, when perfect CSI is available, there is no difference between uncorrelated, correlation length 2 and correlation length 4. However, in the case of imperfect CSI, increase in the correlation length results in worse performance. As could be seen, in uncorrelated, correlation length 2 and 4, the proposed system performs better than conventional system.

### IV- **Conclusion**

According that the optimum detection obtains by optimum estimation, any error in estimation would result in a big fault. This paper, for the first time, presented a deep learning based structure for combating this issue. In order to show the effect of using deep learning, the symbol error rate of a simple FSO communication system is simulated (for the first time) over correlated and un-correlated log-normal channel with perfect/ imperfect CSI. The proposed structure was compared with conventional structures, it was shown that in perfect CSI, both have the same performance, but in imperfect CSI, proposed structure outperforms in channels with un-correlation or desired correlation lengths.